\documentclass[aps,prd,superscriptaddress,nofootinbib,preprintnumbers,floatfix,tightenlines,fleqn,notitlepage,10pt]{revtex4-1}
\usepackage{amsmath,amsfonts,amssymb,amscd,amsxtra,amsthm}
\usepackage{graphicx}  
\usepackage{epstopdf}
\usepackage{dcolumn}  
\usepackage{bm}          
\usepackage{slashed}
\usepackage{cancel}
\usepackage{float}
\usepackage{mathtools} 
\usepackage{amsbsy}
\usepackage{amstext}

\usepackage[utf8]{inputenc}
\usepackage{booktabs}
\usepackage[normalem]{ulem}
\usepackage[dvipsnames]{xcolor}
\usepackage{tabularx}
\usepackage{enumitem}
\usepackage{array}
\usepackage{slashed}
\usepackage{tikz}
\usepackage{pgfplots}
\pgfplotsset{compat=1.18}  
\usepackage{float}
\usepackage{multirow}
\usepackage{cancel}
\usepackage{physics}
\usepackage{hyperref}
\usepackage{cleveref}

\renewcommand\sout{\bgroup \color{red} \ULdepth=-.5ex \ULset}

\begin{document}
\preprint{INHA-NTG-01/2026}
\title{Dynamical generation of charmonium-like tetraquarks
  in an off-shell coupled-channel formalism} 
\author{Hee-Jin Kim}
\affiliation{Institute of Quantum Science, Inha University, Incheon 22212,
  Republic of Korea} 
\author{Hyun-Chul Kim}
\affiliation{Department of Physics, Inha University, Incheon 22212,
  Republic of Korea}
\affiliation{Institute of Quantum Science, Inha University, Incheon 22212,
  Republic of Korea} 
\affiliation{School of Physics, Korea Institute for Advanced Study
(KIAS), Seoul 02455, Republic of Korea}
\date{\today}
\begin{abstract}
We investigate the dynamical generation of charmonium-like ($I=0$)
with spin-parity $J^{PC}=0^{++}, 1^{++}, 2^{++}$, and $3^{--}$ in the
mass range of $3.6$ to $4.3$ GeV. We employ the off-shell
coupled-channel formalism, constructing kernel
amplitudes from effective Lagrangians that respect heavy-quark
spin-flavor and chiral symmetries. To focus solely on dynamically
generated states, we explicitly exclude $s$-channel pole diagrams and
include only $t$- and $u$-channel meson exchanges. Solving the
integral equations, we identify six poles in the complex energy
plane. In the scalar ($0^{++}$) sector, we find a bound state below
the $D\bar{D}$ threshold and a resonance at
$\sqrt{s_R}=(3861-i\,23)\,\mathrm{MeV}$. For the axial-vector
($1^{++}$) sector, the experimentally
observed $\chi_{c1}(3872)$ is reproduced as a bound state near the
$D\bar{D}^*$ threshold, alongside a broader resonance at
$(3961-i\,32)\,\mathrm{MeV}$, which is a plausible candidate for the
$X(3940)$. Furthermore, we find a narrow tensor ($2^{++}$) state at
$4005\,\mathrm{MeV}$ and a vector ($3^{--}$) state at
$4030\,\mathrm{MeV}$. The present results demonstrate that 
coupled-channel dynamics, particularly involving the $D^*\bar{D}^*$
channel, play a crucial role in the formation of these charmonium-like
exotic states. 
\end{abstract}

\maketitle

\section{Introduction}
The discovery of the $\chi_{c1}(3872)$ (also known as $X(3872)$) by
the Belle Collaboration~\cite{Belle:2003nnu} opened a new era in the
study of the heavy hadrons. Its extremely narrow width,
isospin-violating decays, and proximity to the $D^0\bar{D}^{*0}$
threshold were not able to accommodate a simple interpretation within 
conventional quark models (CQM). Since then, a large number of
charmonium-like structures, often referred to as the $XYZ$ states,
have been reported by the Belle, BaBar, BESIII, and LHCb
Collaborations in various hadronic processes such as $B$ decays,
initial-state radiation, two-photon fusion, and $e^+e^-$ annihilation
(see the reviews~\cite{Lebed:2016hpi, Guo:2017jvc, Guo:2019twa,
  Yamaguchi:2019vea, Chen:2022asf, Meng:2022ozq}).
While a few of the observed structures can still be understood as
ordinary charmonia~\cite{Eichten:1978tg,Barnes:2005pb,Li:2009zu},
e.g., $\psi_2(3823)$ and $\psi_3(3842)$ as the $1D$
spin-triplet~\cite{Wang:2015xsa} or $\chi_{c2}(3930)$ as the
$2{}^3P_2$ $c\bar{c}$ state~\cite{Cao:2012du,Wang:2013lpa}, many
others lie close to open-charm thresholds and show behaviors that are
difficult to describe with a naive $c\bar{c}$ interpretation.
Several of these states have unexpectedly narrow widths, decay modes
or branching ratios that do not follow quark-model predictions.

Their properties often depend sensitively on nearby threshold, for
instance $D^{(*)}\bar{D}^{(*)}$ channels, indicating that
coupled-channel dynamics play a crucial role. As a result, the
spectrum in the charmonium sector (especially in the $\sqrt{s}\approx
3.7$ to 
$4.1$~GeV region) cannot be viewed as typical radial or orbital
excitations of $c\bar{c}$ states. Instead, it has become a mixture of
various structures affected by threshold effects and channel
mixing~\cite{Guo:2017jvc,Guo:2019twa}. In such a regime, the
investigation of coupled-channel dynamics provides essential
information: They are directly influenced by how each state couples to
nearby open-flavor channels, such as through threshold effects or
triangle singularities, and therefore offer a practical guide to the
underlying configurations of these hidden-charm candidates.
The theoretical interpretation of the charmonium-like spectrum remains
highly nontrivial, and many models have been developed to
describe the observed structures. Beyond the CQM description, compact
tetraquarks~\cite{Cheng:2020wxa, Praszalowicz:2022sqx, Zhang:2022qtp,
  Kucab:2024nkv, Anwar:2023svj, Meng:2023for, Dong:2024upa} and
hadronic molecules generated by near-threshold
dynamics~\cite{Guo:2017jvc, Guo:2019twa, Yamaguchi:2019vea,
  Meng:2022ozq, Chen:2022asf} constitute two of the most widely
discussed pictures. Additional mechanisms such as hybrid charmonia,
hadroquarkonium configurations, and kinematic effects (cusp-like
structures or triangle singularities) have also been considered in
describing specific candidates. 

A particularly important aspect of this sector is that various
channels are open near $D\bar{D}$, $D\bar{D}^*$, $D^*\bar{D}$, and
$D^*\bar{D}^*$ thresholds, a feature that requires the consideration
of coupled-channel dynamics. While the Bethe-Salpeter (BS)
equation provides a proper theoretical framework for treating multi-channel
hadronic scattering, it is practically too complicated to deal with
the four-dimensional BS equation. To overcome the  
complexities of solving the full BS equation, various approximations  
have been developed such as the on-shell approximation and 
three-dimensional reductions. In the current work, we will take a
three-dimensional reduction known as the Blankenbecler-Sugar
(BbS) scheme, which offers a significant 
advantage by preserving both the unitarity and the off-shell
contributions of the coupled-channel BbS integral equations.
The kernel amplitudes for the BbS equation are constructed by
computing the Feynman invariant amplitudes based on effective
Lagrangians. Since we are mainly interested in how the tetraquark
states are dynamically generated, we only include the $t$- and
$u$-channel amplitudes. To preserve the unitarity and to consider the
structure of hadrons, it is essential to introduce form factors at
each vertex, which causes uncertainties in numerical results. To
minimize them, we introduce the reduced cutoff mass defined by the
subtraction of the corresponding cutoff mass and exchange
hadron mass, which will be discussed in more detail later.
Having solved the three-dimensional coupled BbS
equations, we can derive the transition amplitudes containing the
coupled-channel effects. Surveying pole positions of the transition
amplitudes in the complex energy plane, we can determine the masses
and widths of the tetraquark states corresponding to the poles.
This coupled-channel formalism has been successfully applied to describing
various hadronic reactions.  In Refs.~\cite{Clymton:2022jmv,
  Clymton:2023txd, Clymton:2024pql}, it was shown how the axial-vector
mesons such as $a_1(1260)$, $b_1(1235)$, and $h_1(1415)$ are
dynamically generated from pseudoscalar-vector meson scattering.
The production mechanisms of the hidden-charm pentaquarks
with strangeness varying from $0$ to $-3$ were also investigated within
the same coupled-channel formalism~\cite{Clymton:2024fbf,
  Clymton:2025dzt, Clymton:2025zer}, and the $P_{c\bar{c}}$'s were
interpreted as the molecular states associated with the heavy mesons 
$\bar{D}(\bar{D}^*)$ and singly heavy baryons $\Sigma_c (\Sigma_c^*)$.   
The same theoretical framework was also employed to describe
the $D_{s0}^*(2317)$ as a $DK$ molecular state~\cite{Kim:2023htt} and
to investigate the doubly-charmed tetraquark state
$T_{cc}$~\cite{Kim:2025ado}. A merit of this off-shell coupled-channel
formalism can be found in the fact that it incorporates relevant
symmetries and energy scales across not only $S$ waves but 
also higher partial-wave components. 

\begin{figure}[htp]
  \centering
  \includegraphics{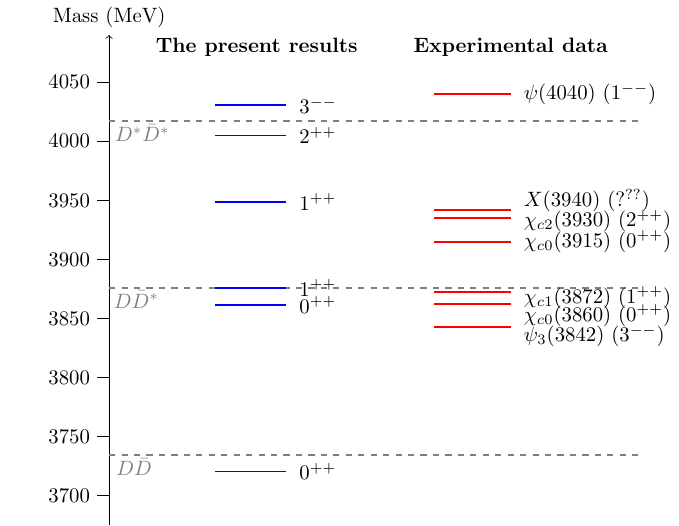}
  \caption{Mass spectrum of hidden-charm states in the range of 
  $3.7$–$4.1$\,GeV. The lines in the left column indicate the
  resonance states obtained in the present work, whereas those in the
  right column denote the experimental data for the states with
  $c\bar{c}$ content~\cite{PDG:2024cfk}.}
  \label{fig:1}
\end{figure}
In the present study, we investigate the dynamical generation of
charmonium-like states ($\chi_{cJ}$ and $\psi_J$ states) with the
hidden-charm $c\bar{c}$ content in the range of
center-of-mass energies from $3.7$ to $4.1$~GeV.
Figure~\ref{fig:1} summarizes the results of this work in
comparison with the conventional $c\bar{c}$ states listed by the
Particle Data Group (PDG)~\cite{PDG:2024cfk}.
Since we mainly focus on the dynamical generation of exotic states, we
intentionally exclude $s$-channel pole diagrams from the kernel
amplitudes to ensure that all resonances are dynamically generated
solely from coupled-channel dynamics. Consequently, we identify six
$c\bar{c}$ states by solving the coupled-channel integral
equations. Three of these states are consistent with the
experimentally observed charmonium-like structures, such as the
$\chi_{c1}(3872)$ near the $D\bar{D}^*$ threshold. Furthermore, our
results predict three additional $c\bar{c}$ resonances that are
dynamically generated through coupled-channel dynamics.

This paper is organized as follows. In Section~II, we explain the
general formalism for the coupled-channel BbS scheme in detail. The
interactions between heavy mesons, charmonia, and light mesons are
derived from effective Lagrangians governed by heavy-quark spin-flavor
and chiral symmetries, which serve as the basis for constructing the
kernel matrices. In Section~III, we discuss the numerical results for
each spin-parity assignment. By scanning the transition amplitudes, we
extract pole positions and coupling strengths to various hadronic
channels to elucidate their dynamical origins and properties.
We will also discuss the physical origin of the agreement and
discrepancy with the experimental data in detail. Section~IV is
devoted to a summary of our numerical analysis and concluding
remarks. 

\section{General Formalism}
\begin{figure}[htp]
  \centering
  \includegraphics{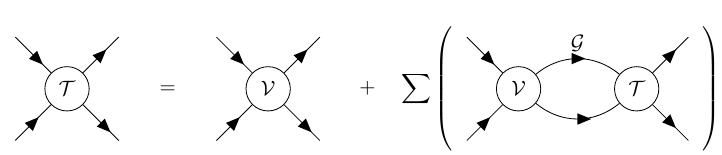}
  \caption{Schematic diagram for the two-body coupled integral
    equations.} 
  \label{fig:2}
\end{figure}

The scattering amplitude of the $i\to f$ process is given by
\begin{align}
  S_{fi} = \delta_{fi} - (2\pi)^4 \delta^{(4)}(P_f-P_i) \mathcal{T}_{fi},
\label{eq:1}
\end{align}
where $P_i$ and $P_f$ are the total four-momenta of the initial and
final states, respectively. The transition amplitudes $\mathcal{T}_{fi}$
contain nontrivial information on how the resonances and bound states
are generated dynamically. The $\mathcal{T}_{fi}$ are obtained by
solving the Bethe-Salpeter (BS) equation with the two-body Feynman
kernel amplitudes:
\begin{align}
  \mathcal{T}_{fi}(p,p';s) &= \mathcal{V}_{fi}(p,p';s)
  \frac{1}{(2\pi)^4} \sum_k \int d^4q_k\, 
  \mathcal{V}_{fk}(p,q_k;s) G_k(q_k;s) \mathcal{T}_{ki}(q_k,p';s),
\label{eq:2} 
\end{align}
where $p$ and $p'$ denote respectively the relative four-momenta of 
the initial and final states, and $q_k$ is the off-shell four-momentum
for intermediate states in the center-of-mass (CM) frame. The variable 
$s \equiv P_i^2 = P_f^2$ is the square of the total energy. The indices 
$i$, $f$, and $k$ indicate the initial, final, and intermediate states, 
respectively. The index $k$ runs over all relevant intermediate channels 
to nonstrange hidden-charm states. The schematic diagram for the coupled 
BS equation~\eqref{eq:2} is depicted in Fig.~\ref{fig:2}.  

Since it is computationally demanding to solve the four-dimensional
BS integral equations, we introduce a three-dimensional reduction
approach. While there are various ways for this reduction, we adopt 
the Blankenbecler-Sugar formalism (BbS)~\cite{Blankenbecler:1965gx,
  Aaron:1968aoz}, where the two-body propagator is taken to be   
\begin{align}
  G_k(q_k) &= \frac{\pi}{\omega_1^k \omega_2^k} \delta\left(q_k^0 - 
  \frac{\omega_1^k-\omega_2^k}{2}\right) 
  \frac{\omega_1^k+\omega_2^k}{s-(\omega_1^k+\omega_2^k)^2 +i\epsilon},  
\label{eq:3}
\end{align}
where $\omega_{1(2)}^k=\sqrt{\bm{q}_k^2+(m_{1(2)}^k)^2}$ is the energy
of the particle $1(2)$ in the channel $k$ on the mass shell.
We then insert $G_k$ in Eq.~\eqref{eq:2}, so that we obtain 
the three-dimensional coupled integral equation for each $q_k^0$:
\begin{align}
  T_{fi}(\bm{p},\bm{p}';s) &= V_{fi}(\bm{p},\bm{p}';s) 
  + \frac{1}{(2\pi)^3} \sum_k \int \frac{d^3 q_k}{2\omega_1^k \omega_2^k}
  \frac{\omega_1^k + \omega_2^k}{s-(\omega_1^k+\omega_2^k)^2+i\epsilon} 
  V_{fk}(\bm{p},\bm{q}_k;s) T_{ki}(\bm{q}_k,\bm{p}';s). 
\label{eq:4}
\end{align}
To investigate dynamically generated resonances with specific quantum
numbers, we need to perform the partial-wave decomposition of the
transition amplitudes as follows: 
\begin{align}
  T_{\lambda'\lambda}^{J(fi)} (\mathrm{p},\mathrm{p}';s) 
  &= V_{\lambda'\lambda}^{J(fi)}(\mathrm{p},\mathrm{p}';s) 
  + \sum_{k,\{\lambda_k\}} \frac{1}{(2\pi)^3} \int_0^\infty d\mathrm{q}_k\, 
  \frac{\omega_1^k+\omega_2^k}{2\omega_1^k\omega_2^k}
  \frac{\mathrm{q}_k^2 V_{\lambda'\lambda_k}^{J(fk)}(\mathrm{p},\mathrm{q}_k;s) 
  T_{\lambda_k\lambda}^{J(ki)}(\mathrm{q}_k,\mathrm{p}';s)}
  {s-(\omega_1^k+\omega_2^k)^2+i\epsilon}, 
\label{eq:5}
\end{align}
where $\{\lambda_k\}$ stands for the helicity components for the
particles involved in the intermediate hadron channel $k$. The
partial-wave kernel and transition amplitudes in the helicity basis are
derived as  
\begin{align}
  V_{\lambda'\lambda}^{J(fi)}(\mathrm{p},\mathrm{p}';s) 
  &= 2\pi \int d\cos\theta \, V_{fi}(\mathrm{p},\mathrm{p}',\theta;s)  
  d_{\lambda\lambda'}^J(\theta), \\
  T_{\lambda'\lambda}^{J(fi)}(\mathrm{p},\mathrm{p}';s) 
  &= 2\pi \int d\cos\theta \, T_{fi}(\mathrm{p},\mathrm{p}', \theta;s)
  d_{\lambda\lambda'}^J(\theta),
\label{eq:6}
\end{align}
where $d_{\lambda\lambda'}^J$ denotes the reduced Wigner $D$ function
as a function of the scattering angle $\theta$. $\lambda^{(')}$
indicates the helicity difference between two particles in the initial
(final) state. $\mathrm{p}$, $\mathrm{p}'$, and $\mathrm{q}_k$ are
the magnitudes of three-momenta for initial, final, and intermediate
states, respectively.

To solve the coupled integral equation~\eqref{eq:5}, we first
construct the kernel matrix $V_{fi}$ based on the effective
Lagrangians~\cite{Wise:1992hn,Yan:1992gz,Casalbuoni:1996pg} with
heavy-quark spin-flavor symmetry and chiral symmetry, given as  
\begin{align} 
  \mathcal{L}_\mathrm{heavy}^Q 
    &= ig \mathrm{Tr} [H_b^Q \gamma_\mu \gamma_5 \mathcal{A}_{ba}^\mu 
    \bar{H}_a^Q] + i\beta \mathrm{Tr} [H_b^Q v_\mu (\mathcal{V} - 
    \rho)_{ba}^\mu \bar{H}_a^Q] + i\lambda \mathrm{Tr} [H_b^Q \sigma_{\mu\nu} 
    F_{ba}^{\mu\nu} \bar{H}_a^Q] + g_\sigma \bar{H}_a^Q H_a^Q \sigma, \\
  \mathcal{L}_\mathrm{heavy}^{\bar{Q}}
    &= ig \mathrm{Tr} [\bar{H}_a^{\bar{Q}} \gamma_\mu \gamma_5 
    \mathcal{A}_{ab}^\mu H_b^{\bar{Q}}] - i\beta \mathrm{Tr} 
    [\bar{H}_a^{\bar{Q}} v_\mu (\mathcal{V}-\rho)_{ab}^\mu 
    H_b^{\bar{Q}}] + i\lambda \mathrm{Tr} [\bar{H}_a^{\bar{Q}} \sigma_{\mu\nu} 
    F_{ab}^{\mu\nu} H_b^{\bar{Q}}] + g_\sigma \bar{H}_a^{\bar{Q}} 
    H_a^{\bar{Q}} \sigma, 
\label{eq:7}
\end{align}
where $H^Q$ and its conjugate field $\bar{H}^Q$ represent the heavy
superfields that contain a heavy quark $Q$, and $H^{\bar{Q}}$ and
$\bar{H}^{\bar{Q}}$ denote those for superfields with anti-heavy quark
$\bar{Q}$. The heavy-quark spin-flavor symmetry allows us to write
down the field $H$ in terms of the pseudoscalar and vector heavy-meson
fields as follows: 
\begin{align} 
  H_a^Q 
    &= \frac{1+\slashed{v}}{2}(P_a^{*Q\mu} \gamma_\mu - 
    P_a^Q \gamma_5), \nonumber \\
  \bar{H}_a^Q 
    &= \gamma_0 H_a^{Q\dagger} \gamma_0
    = (P_a^{*Q\dagger\mu} \gamma_\mu + P_a^{Q\dagger} \gamma_5) 
    \frac{1+\slashed{v}}{2}, \\
  H_a^{\bar{Q}} 
    &= (P_a^{*\bar{Q}\mu} \gamma_\mu - P_a^{\bar{Q}\mu} \gamma_5) 
    \frac{1-\slashed{v}}{2}, \nonumber \\
  \bar{H}_a^{\bar{Q}} 
    &= \gamma^0 H_a^{\bar{Q}\dagger} \gamma^0
    = \frac{1-\slashed{v}}{2} \left[ \slashed{P}_a^{*\bar{Q}} + 
    P_a^{\bar{Q}} \gamma_5 \right],
\label{eq:8}
\end{align}
where $P^Q$ and $P^{*Q}$ stand for the heavy meson antitriplet in
flavor SU(3) symmetry with a heavy quark $Q$ regarded as the
static color source. These fields are normalized by 
\begin{align} 
  \langle 0 | P | D(B) \rangle &= \sqrt{M_{D(B)}}, \\
  \langle 0 | P^{*\mu} | D^*(B^*) \rangle &= 
  \epsilon^\mu \sqrt{M_{D^*(B^*)}},
\label{eq:9}
\end{align}
where $\epsilon^\mu$ designates the polarization vector of the heavy 
vector meson. The SU(3) triplet field $P^{\bar{Q}}$ and $P^{*\bar{Q}}$
are also defined in the same manner. The interactions of the
pseudo-Nambu-Goldstone (pNG) bosons with the matter field $H$ can be
derived by the non-linear realization of the pNG fields. The coset
field $\xi(x)$, which arises from spontaneous breakdown of chiral
symmetry, is parametrized in the coset space $\mathrm{SU(3)}_L \otimes
\mathrm{SU(3)}_R / \mathrm{SU(3)}_V$ as $\xi(x) = \exp(i\mathcal{M}/
f_\pi)$ with the pion decay constant $f_\pi = 132$ MeV taken as a 
normalization factor. Here the $\mathcal{M}$ represents the matrix in 
flavor space that describes the SU(3) pNG fields or the light 
pseudoscalar meson matrix:
\begin{align} 
  \mathcal{M} = \left(
  \begin{array}{ccc}
    \frac{\pi^0}{\sqrt{2}} + \frac{\eta}{\sqrt{6}} & \pi^+ & K^+ \\ 
    \pi^- & -\frac{\pi^0}{\sqrt{2}} + \frac{\eta}{\sqrt{6}} & K^0 \\
    K^- & \bar{K}^0 & -\sqrt{\frac{2}{3}}\eta
  \end{array} \right).
\label{eq:10}
\end{align}
One can introduce the axial-vector and vector currents:
\begin{align} 
  \mathcal{A}^\mu &= \frac{1}{2} (\xi^\dagger \partial^\mu \xi - 
  \xi \partial^\mu \xi^\dagger), \\
  \mathcal{V}^\mu &= \frac{1}{2} (\xi^\dagger \partial^\mu \xi + 
  \xi \partial^\mu \xi^\dagger),
\label{eq:11}
\end{align}
and the axial-vector current can be expanded as:
\begin{align} 
  \mathcal{A}^\mu &= \frac{1}{2} (\xi^\dagger \partial^\mu \xi - 
  \xi \partial^\mu \xi^\dagger) 
  = \frac{i}{f_\pi} \partial^\mu \mathcal{M} + \cdots.
\label{eq:12}
\end{align}
It indicates that the derivative coupling is natural for the
light pseudoscalar meson and heavy mesons on account of chiral
symmetry and its spontaneous breakdown. 

The effective Lagrangian for the vector meson octet is constructed
by means of the hidden local symmetry approach. Bando et al.~
\cite{Wise:1992hn,Bando:1985rf} suggested the gauge equivalence
between $\mathrm{SU(3)}_L \otimes \mathrm{SU(3)}_R / \mathrm{SU(3)}_V$
and $[\mathrm{SU(3)}_L \otimes \mathrm{SU(3)}_R] \otimes [\mathrm{SU
(3)}_V]_\mathrm{local}$, where $[\mathrm{SU(3)}_V]_\mathrm{local}$
stands for the hidden local symmetry and the dynamical gauge bosons
appear as the composite fields. The invariance under the local
symmetry requires the new interaction terms in the effective
Lagrangian corresponding to the third and fourth term in Eq.~
\eqref{eq:7}. The vector meson octet is then expressed as:
\begin{align} 
  \rho^\mu &= i\frac{g_V}{\sqrt{2}} V^\mu, \;\;\;
  V^\mu = \left( \begin{array}{ccc}
    \frac{\rho^0}{\sqrt{2}} + \frac{\omega}{\sqrt{6}} & 
    \rho^+ & K^{*+} \\ 
    \rho^- & -\frac{\rho^0}{\sqrt{2}} + \frac{\omega}{\sqrt{6}} & 
    K^{*0} \\
    K^{*-} & \bar{K}^{*0} & -\sqrt{\frac{2}{3}}\omega
  \end{array} \right)^\mu,
\label{eq:13}
\end{align}
with the field strength tensor $F^{\mu\nu} = \partial^\mu \rho^\nu -
\partial^\nu \rho^\mu + [\rho^\mu, \rho^\nu]$.

The expansion of the effective Lagrangian in Eq.~\eqref{eq:7}
represents the interaction terms for $P^{Q(\bar{Q})}$,
$P^{*Q(\bar{Q})}$, $\mathcal{M}$, and $V$. The first term of
Eq.~\eqref{eq:1} gives the effective Lagrangians for $PP^*\mathcal{M}$
and $P^*P^*\mathcal{M}$ vertices:
\begin{align} 
  \mathcal{L}_{P^Q P^{*Q} \mathcal{M}} 
    &= -\frac{2g}{f_\pi} P_b^{*Q\mu} \partial_\mu \mathcal{M}_{ba} 
    P_a^{Q\dagger} + \mathrm{h.c.}, \\
  \mathcal{L}_{P^{*Q} P^{*Q} \mathcal{M}} 
    &= \frac{2ig}{f_\pi} P_b^{*Q\beta} \partial^\mu \mathcal{M}_{ba} 
    v^\nu P_a^{*Q\alpha\dagger} \varepsilon_{\alpha\beta\mu\nu} 
    + \mathrm{h.c.}, \\
  \mathcal{L}_{P^{\bar{Q}} P^{*\bar{Q}} \mathcal{M}} 
    &= \frac{2g}{f_\pi} P_b^{*\bar{Q}\mu} \partial_\mu 
    \mathcal{M}_{ba} P_a^{\bar{Q}\dagger} + \mathrm{h.c.}, \\
  \mathcal{L}_{P^{*\bar{Q}} P^{*\bar{Q}} \mathcal{M}} 
    &= \frac{2ig}{f_\pi} P_b^{*\bar{Q}\beta} \partial^\mu 
    \mathcal{M}_{ba} v^\nu P_a^{*\bar{Q}\alpha\dagger} 
    \varepsilon_{\alpha\beta\mu\nu} + \mathrm{h.c.}.
\label{eq:14}
\end{align}

The interactions of $PPV$, $PP^*V$, and $P^*P^*V$ then come from the
second and third terms of Eq.~\eqref{eq:7}:
\begin{align} 
  \mathcal{L}_{P^Q P^Q V} 
    &= -2i\beta P^Q P^{Q\dagger} v \cdot \rho \nonumber \\
    &= -\sqrt{2}\beta g_V P^Q P^{Q\dagger} v \cdot V, \\
  \mathcal{L}_{P^Q P^{*Q} V} 
    &= -4i\lambda \varepsilon_{\mu\nu\alpha\beta} (P^Q \partial^\mu 
    P^{*Q\dagger\nu} + P^{Q\dagger} \partial^\mu P^{*Q\nu}) 
    v^\alpha \rho^\beta \nonumber \\
    &= 2\sqrt{2}\lambda g_V \varepsilon_{\mu\nu\alpha\beta} 
    (P^Q \partial^\mu P^{*Q\dagger\nu} + P^{Q\dagger} \partial^\mu 
    P^{*Q\nu}) v^\alpha V^\beta, \\
  \mathcal{L}_{P^{*Q} P^{*Q} V} 
    &= -2i\beta P^{*Q\mu} P_\mu^{*Q\dagger} v \cdot \rho 
    - 4\lambda (\partial_\mu P^{*Q\dagger\mu} P_\nu^{*Q} 
    - \partial_\mu P^{*Q\mu} P_\nu^{*Q\dagger}) \rho^\nu \nonumber \\
    &= \sqrt{2}\beta g_V P^{*Q\mu} P_\mu^{*Q\dagger} v \cdot V 
    - 2\sqrt{2}i\lambda g_V (\partial_\mu P^{*Q\dagger\mu} 
    P_\nu^{*Q} - \partial_\mu P^{*Q\mu} P_\nu^{*Q\dagger}) V^\nu, \\
  \mathcal{L}_{P^{\bar{Q}} P^{\bar{Q}} V} 
    &= 2i\beta P^{\bar{Q}} P^{\bar{Q}\dagger} v \cdot \rho \nonumber \\
    &= \sqrt{2}\beta g_V P^{\bar{Q}} P^{\bar{Q}^\dagger} v \cdot V, \\
  \mathcal{L}_{P^{\bar{Q}} P^{*\bar{Q}} V} 
    &= -4i\lambda \varepsilon_{\mu\nu\alpha\beta} (P^{\bar{Q}} 
    \partial^\mu P^{*\bar{Q}\dagger\nu} + P^{\bar{Q}\dagger} 
    \partial^\mu P^{*\bar{Q}\nu}) v^\alpha \rho^\beta \nonumber \\
    &= 2\sqrt{2}\lambda g_V \varepsilon_{\mu\nu\alpha\beta} 
    (P^{\bar{Q}} \partial^\mu P^{*\bar{Q}\dagger\nu} + 
    P^{\bar{Q}\dagger} \partial^\mu P^{*\bar{Q}\nu}) 
    v^\alpha V^\beta, \\
  \mathcal{L}_{P^{*\bar{Q}} P^{*\bar{Q}} V} 
    &= -2i\beta P^{*\bar{Q}\mu} P_\mu^{*\bar{Q}\dagger} v \cdot \rho 
    + 4\lambda (\partial_\mu P^{*\bar{Q}\dagger\mu} P_\nu^{*\bar{Q}} 
    - \partial_\mu P^{*\bar{Q}\mu} P_\nu^{*\bar{Q}\dagger}) 
    \rho^\nu \nonumber \\
    &= \sqrt{2}\beta g_V P^{*\bar{Q}\mu} P_\mu^{*\bar{Q}\dagger} 
    v \cdot V + 2\sqrt{2}i\lambda g_V (\partial_\mu 
    P^{*\bar{Q}\dagger\mu} P_\nu^{*\bar{Q}} - \partial_\mu 
    P^{*\bar{Q}\mu} P_\nu^{*\bar{Q}\dagger}) V^\nu.
\label{eq:15}
\end{align}

The effective Lagrangians for interaction with the scalar meson
$\sigma$ are obtained as:
\begin{align}
  \mathcal{L}_{P^Q P^Q \sigma} 
    &= -2g_{P^Q P^Q \sigma} \, P_a^{Q\dagger} P_a^Q \, \sigma, \\
  \mathcal{L}_{P^{*Q} P^{*Q} \sigma} 
    &= 2g_{P^{*Q} P^{*Q} \sigma} \, P_{a\mu}^{*Q\dagger} 
    P_a^{*Q\mu} \, \sigma, \\
  \mathcal{L}_{P^{\bar{Q}} P^{\bar{Q}} \sigma} 
    &= -2g_{P^{\bar{Q}} P^{\bar{Q}} \sigma} \, P_a^{\bar{Q}\dagger} 
    P_a^{\bar{Q}} \, \sigma, \\
  \mathcal{L}_{P^{*\bar{Q}} P^{*\bar{Q}} \sigma} 
    &= 2g_{P^{*\bar{Q}} P^{*\bar{Q}} \sigma} \, 
    P_{a\mu}^{*\bar{Q}\dagger} P_a^{*\bar{Q}\mu} \, \sigma,
\label{eq:16}  
\end{align}
where the heavy quark velocity can be replaced by
$i\overleftrightarrow{\partial}_\mu/(2\sqrt{MM'}) =
i(\overleftarrow{\partial}-\overrightarrow{\partial})/(2\sqrt{MM'})$,
with $M^{(')}$ denoting the mass of the heavy meson.

Charmonia and light unflavored mesons can be involved in the energy
region, where hidden-charm exotic mesons are dynamically
generated. The corresponding effective Lagrangian is given by:
\begin{align}
  \mathcal{L}_J = g_\psi \mathrm{Tr} [J \bar{H}_a^{\bar{Q}} \gamma_\mu 
  \partial^\mu \bar{H}_a^Q],
\label{eq:17}
\end{align}
where $J$ represents the charmonium superfield containing both
pseudoscalar ($\eta_c$) and vector ($J/\psi$) components:
\begin{align}
  J = \frac{1+\slashed{v}}{2} \left[ \psi^\mu \gamma_\mu - 
  \eta_c \gamma_5 \right] \frac{1-\slashed{v}}{2}.
\label{eq:18}
\end{align}
We get the corresponding effective Lagrangian for the pseudoscalar
charmonium $\eta_c$:   
\begin{align}
  \mathcal{L}_{\eta_c} 
    &= -g_\psi \eta_c (P^{\bar{Q}\dagger} v \cdot P^{*Q\dagger} + 
    P^{Q\dagger} v \cdot P^{*\bar{Q}\dagger}) 
    -ig_\psi \eta_c \varepsilon_{\beta\alpha\mu\nu} 
    P^{*\bar{Q}\alpha\dagger} v^\beta P^{*Q\dagger\mu} v^\nu,
\label{eq:19}
\end{align}
and for the vector charmonium $J/\psi$ $(\psi^\mu)$: 
\begin{align}
  \mathcal{L}_{J/\psi} 
    &= ig_\psi \varepsilon^{\beta\alpha\mu\nu} v_\nu \psi_\alpha 
    \left( P^{Q\dagger} v_\beta P_\mu^{*\bar{Q}\dagger} + 
    P_\beta^{*Q\dagger} v_\alpha P^{\bar{Q}\dagger} \right)
    + \psi_\mu P_\nu^{*Q\dagger} v_\alpha P_\beta^{*\bar{Q}\dagger} 
    (g^{\mu\nu}g^{\alpha\beta} - g^{\mu\alpha}g^{\nu\beta} + 
    g^{\mu\beta}g^{\nu\alpha}) \nonumber \\
    &\quad - g_\psi P^{Q\dagger} v \cdot \psi P^{\bar{Q}\dagger}.
\label{eq:20}
\end{align}

We now discuss how the values of the coupling constants in
Eq.~\eqref{eq:7} can be determined. $g$ is the the axial coupling
constant and its value is extracted to be $g=0.59$ from the $D^*\to
D\pi$ decay rate~\cite{CLEO:2001foe}:
\begin{align}
\Gamma(D^{*+} \to D^0\pi^+) = \frac{g^2}{6\pi f_\pi^2} |\bm{p}_\pi|^3,  
\label{eq:21}
\end{align}
The interactions of $PPV$, $PP^*V$, and $P^*P^*V$ then come from the
second and third terms of Eq.~\eqref{eq:7}. $\beta=0.9$ is determined
by the vector meson dominance~ \cite{Bando:1985rf}. The value of 
$\lambda=0.56\, \mathrm{GeV}^{-1}$ is taken from
Ref.~\cite{Isola:2003fh}, which is 
the coupling constant for the third term given in Eq.~\eqref{eq:7}.
Note that it is related to the $PP^*V$ vertex and $P^*P^*V$ tensor 
coupling.  The value of $\lambda$ can be determined from the $B\to
K^*$ vector transition form factor at very high $q^2$ based on the
effective field theory approach. With the results from the light-cone
QCD sum rules and lattice QCD matched, its value is set to be
$\lambda=0.56\,\mathrm{GeV}^{-1}$. The $\sigma$ state (also known as
$f(500)$) in Eq.~\eqref{eq:16} is often regarded as a chiral
partner of the pion, so that the following relation was suggested:
$g_\sigma=g_\pi/(2\sqrt{6})$. In Ref.~\cite{Kim:2019rud}, however, the
large width of the $\sigma$ from dynamics of the correlated $2\pi$
exchange were considered and the values of $g_{DD\sigma}=1.50$ and
$g_{D^*D^*\sigma}5.21$ were extracted by using the disperstion
relation. The $g_{DD\rho}=1.65$ and $g_{D^*D^*\rho}=6.47$ were also  
obtained in the same manner by analyzing the vector-isovector
$2\pi$-correlated exchange. The $g_{DD^*\rho}$ coupling constant is
fixed by using the KSRF relation~\cite{Kawarabayashi:1966kd,
  Riazuddin:1966sw}: $g_{DD^*\rho}=5.8$.  

We now construct the kernel matrix for charmonium-like states,
$\chi_{cJ}$ and $\psi_J$, within the energy range of the $D\bar{D}$,
$D\bar{D}^*$, and $D^*\bar{D}^*$ thresholds. We are able to list the
two-hadron state with $I=0$ lying near these thresholds as follows:
\begin{itemize}
\item $D\bar{D}$, $D\bar{D}^*$, $D^*\bar{D}^*$,
$D_s\bar{D}_s$, $D_s\bar{D}_s^*$, $D_s^*\bar{D}_s^*$
\;\; ($G=\pm1$), 
\item $J/\psi \omega$, $J/\psi \phi$, \;\; ($G=+1$)
\item $\eta_c \omega$, $\eta_c \phi$, \;\; ($G=-1$)
\end{itemize}

To properly define the allowed hadronic channels for the $\chi_{cJ}$
and $\psi_J$ families, we must classify the two-body states according
to their exact quantum numbers. Since our scope is restricted to
the isoscalar sector ($I=0$), the $G$-parity is equivalent to the
charge-conjugation parity, $G = C(-1)^I = C$. Furthermore, because
the constituent mesons in the present study are either pseudoscalar
($J^P=0^-$) or vector ($J^P=1^-$), the intrinsic parity of any
two-meson channel is always positive ($P_1 P_2 = (-1)^2 = +1$).
Consequently, the total parity of the system is determined entirely
by the relative orbital angular momentum $L$, yielding $P = (-1)^L$.

The assignment of $C$-parity requires careful treatment depending on
the nature of the interacting mesons. For hidden-charm channels
consisting of a charmonium and a light vector meson (such as
$J/\psi\,\omega$ and $\eta_c\,\phi$), the total $C$-parity is simply
the product of the intrinsic $C$-parities of the constituents,
$C = C_1 C_2$, completely independent of the relative orbital angular
momentum. For open-flavor channels composed of identical
meson--antimeson pairs (such as $D\bar{D}$ and $D^*\bar{D}^*$), the
$C$-parity depends on both the orbital angular momentum $L$ and the
total spin $S$ of the two-meson system, given by $C = (-1)^{L+S}$.

On the other hand, non-identical open-flavor meson pairs, such as
$D\bar{D}^*$, are not $C$-eigenstates. The $C$-parity eigenstates are
constructed via symmetric or antisymmetric linear combinations:
\begin{align}
|D\bar{D}^*\rangle_{C=\pm} = \frac{1}{\sqrt{2}} \left(
|D\bar{D}^*\rangle \mp |\bar{D}D^*\rangle \right),
\label{eq:22}
\end{align}
with $C|D(\bar D)\rangle = +|D(\bar D)\rangle$ and $C|D^*(\bar
D^*)\rangle= -|D^*(\bar D^*)\rangle$. By imposing these exact
selection rules, we systematically construct two independent kernel
matrices composed of the relevant hadronic channels for the
positive-$G$-parity ($\chi_{cJ}$) and negative-$G$-parity ($\psi_J$)
sectors.

\begin{figure}[htp]
\centering
\includegraphics[scale=.8]{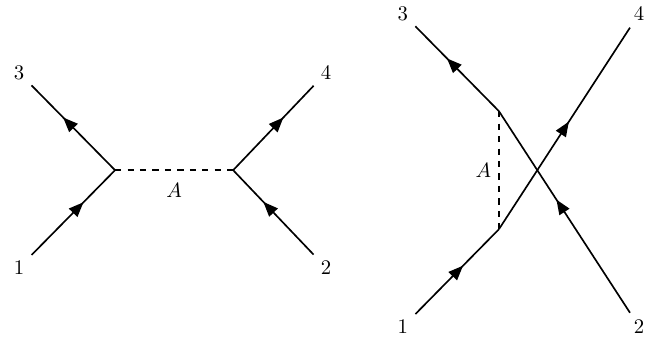}
\caption{$t$- and $u$-channel Feynman diagrams for $12\to 34$.}
\label{fig:3}
\end{figure}

The kernel matrix element is written as the sum of tree-level Feynman
amplitudes for all allowed single-meson exchanges,
\begin{align}
\mathcal{V}_{i\to j} = \sum_A \mathcal{M}_{i\to j}^A ,
\label{eq:23}
\end{align}
where $A$ labels the exchanged particle in the process $i\to j$. As
mentioned in the Introduction, we have intentionally excluded any
$s$-channel contributions in this study, since we focus on the
resonances generated by the coupled-channel formalism.
Figure~\ref{fig:3} shows a generic tree-level diagram with two
vertices and a meson propagator.
The Feynman amplitudes for the exchange of $A$ read
\begin{align}
\mathcal{M}_{12\to 34}^A(t) &= \mathrm{IS}\, F_A^2\, \Gamma_{13}^A\,
P^A\, \Gamma_{24}^A , \cr
\mathcal{M}_{12\to 34}^A(u) &= \mathrm{IS}\, F_A^2\, \Gamma_{14}^A\,
P^A\, \Gamma_{23}^A ,
\label{eq:24}
\end{align}
where $\Gamma_{ij}^A$ denotes the vertex function that couples the $i$
and $j$ legs to the exchanged particle $A$. The vertex functions are
obtained straightforwardly from the effective Lagrangians in
Eqs.~\eqref{eq:14}--\eqref{eq:16} and \eqref{eq:21}. The quantity
$P^A$ is the propagator of the exchanged particle $A$. The propagators
for spin-$0$ and spin-$1$ particles are given by
\begin{align}
P^A(q) &= \frac{1}{q^2 - m_A^2} , \;\;\;
P_{\mu\nu}^A(q) = \frac{-g_{\mu\nu} + \frac{q_\mu q_\nu}{m_A^2}}
{\,q^2 - m_A^2\,} .
\label{eq:25}
\end{align}
The factor $\mathrm{IS}$ denotes the isospin factor arising from the
isospin projector and SU(3) Clebsch--Gordan coefficients.

We introduce a form factor at each vertex, since hadrons have finite
sizes. This form factor is also necessary to ensure the unitarity of
the transition amplitudes. We use the following
form~\cite{Kim:1994ce}:
\begin{align}
F(q^2) = \left(\frac{n\Lambda^2-m_A^2}
{n\Lambda^2-q^2} \right)^n,
\label{eq:42}
\end{align}
where the value of $n$ depends on the momentum power present
in the vertex function $\Gamma$. Note that we have turned off the
energy dependence in Eq.~\eqref{eq:42}, which causes unphysical
behaviors of the kernel amplitudes. 
Although there is no experimental information on the values of the
cutoff masses $\Lambda$ in Eq.~\eqref{eq:42}, one can reduce the
uncertainties associated with them by considering physical
properties of the hadrons involved. Investigations of the
electromagnetic form factors of singly heavy
baryons~\cite{Kim:2018nqf, Kim:2021xpp} have shown that sizes of heavy
hadrons are more compact than those of lighter ones. This
implies that higher cutoff masses are more suitable for heavy hadrons
than for light ones, since the value of the cutoff mass is
proportional to the inverse of the size of the corresponding
hadron. Thus, we define the reduced cutoff mass as 
$\Lambda_0 := \Lambda - m_A$. Interestingly, in
Ref.~\cite{Cheng:2004ru}, the cutoff mass is related to
$\Lambda_{\mathrm{QCD}}$ via $\Lambda = m + \eta
\Lambda_{\mathrm{QCD}}$, where $\eta$ is of
order unity. This method has been successfully applied in previous
works~\cite{Clymton:2022jmv,
  Clymton:2023txd, Kim:2023htt, Clymton:2024pql, Kim:2025ado,
  Clymton:2024fbf, Clymton:2025hez, Clymton:2025zer}.
In the present study, we fix the values of the reduced cutoff to be
$\Lambda_0 = 600$~MeV for all vertices throughout the
calculation. We emphasize that no fitting procedure is employed in the
present study. The values of the isospin factors $\mathrm{IS}$ and the
adopted $\Lambda_0$ for the isoscalar and isovector sectors are listed
in Tables~\ref{tab:1} and \ref{tab:2}, respectively. 

\setlength{\tabcolsep}{15pt}
\renewcommand{\arraystretch}{1.5}
\begin{table}[htp]
\centering
\caption{The values of the SU(3) symmetric factor for the transitions
between open-flavor channels and isospin factor for the kernel
amplitude in isoscalar ($I=0$) channel. $\Lambda_0$ denotes the
reduced cutoff mass defined as $\Lambda-m_\mathrm{ex}$.}
\label{tab:1}
\begin{tabular}{l|c||c|c|c}
\hline
\hline
\multirow{2}{*}{Reaction} & \multirow{2}{*}{Exchange} & 
\multicolumn{2}{c|}{IS factor} & \multirow{2}{*}{$\Lambda_0$ [MeV]}\\  
\cline{3-4}
 & & $t$-ch & $u$-ch & \\
\hline
\multirow[t]{3}{*}{$D\bar{D}\to D\bar{D}$} 
& $\sigma$ & $1$ & $-1$ & $600$ \\
& $\rho$ & $3/2$ & $-3/2$ & $600$ \\
& $\omega$ & $1/2$ & $-1/2$ & $600$ \\
\multirow[t]{2}{*}{$D\bar{D}\to D\bar{D}^*$} 
& $\rho$ & $3/2$ & $-3/2$ & $600$ \\
& $\omega$ & $1/2$ & $-1/2$ & $600$ \\
\multirow[t]{1}{*}{$D\bar{D}\to D_s\bar{D}_s$}
& $K^*$ & $-\sqrt{2}$ & $-\sqrt{2}$ & $600$ \\
\multirow[t]{2}{*}{$D\bar{D}\to D_s\bar{D}_s^*$}
& $K^*$ & $-\sqrt{2}$ & $-\sqrt{2}$ & $600$ \\
\multirow[t]{4}{*}{$D\bar{D}\to D^*\bar{D}^*$} 
& $\pi$, $\rho$ & $3/2$ & $-3/2$ & $600$ \\
& $\eta$ & $1/6$ & $-1/6$ & $600$ \\
& $\omega$ & $1/2$ & $-1/2$ & $600$ \\
\multirow[t]{2}{*}{$D\bar{D}\to D_s^*\bar{D}_s^*$}
& $K$, $K^*$ & $-\sqrt{2}$ & $-\sqrt{2}$ & $600$ \\
\hline
\multirow[t]{5}{*}{$D\bar{D}^*\to D\bar{D}^*$} 
& $\sigma$ & $1$ & - & $600$ \\
& $\pi$ & - & $-3/2$ & $600$ \\
& $\eta$ & - & $1/6$ & $600$ \\
& $\rho$ & $3/2$ & $-3/2$ & $600$ \\
& $\omega$ & $1/2$ & $-1/2$ & $600$ \\
\multirow[t]{2}{*}{$D\bar{D}^*\to D_s\bar{D}_s$}
& $K^*$ & $-\sqrt{2}$ & $-\sqrt{2}$ & $600$ \\
\multirow[t]{4}{*}{$D\bar{D}^*\to D^*\bar{D}^*$} 
& $\pi$, $\rho$ & $3/2$ & $-3/2$ & $600$ \\
& $\eta$ & $1/6$ & $-1/6$ & $600$ \\
& $\omega$ & $1/2$ & $-1/2$ & $600$ \\
\multirow[t]{2}{*}{$D\bar{D}^*\to D_s\bar{D}_s^*$}
& $K$ & - & $-\sqrt{2}$ & $600$ \\
& $K^*$ & $-\sqrt{2}$ & $-\sqrt{2}$ & $600$ \\
\multirow[t]{2}{*}{$D\bar{D}^*\to D_s^*\bar{D}_s^*$}
& $K$, $K^*$ & $-\sqrt{2}$ & $-\sqrt{2}$ & $600$ \\
\hline
\multirow[t]{2}{*}{$D_s\bar{D}_s\to D_s\bar{D}_s$}
& $\sigma$, $\phi$ & $1$ & $1$ & $600$ \\
\multirow[t]{2}{*}{$D_s\bar{D}_s\to D^*\bar{D}^*$}
& $K$, $K^*$ & $-\sqrt{2}$ & $-\sqrt{2}$ & $600$ \\
\multirow[t]{2}{*}{$D_s\bar{D}_s\to D_s\bar{D}_s^*$}
& $\phi$ & $1$ & $1$ & $600$ \\
\multirow[t]{2}{*}{$D_s\bar{D}_s\to D_s^*\bar{D}_s^*$}
& $\eta$ & $2/3$ & $2/3$ & $600$ \\
& $\phi$ & $1$ & $1$ & $600$ \\
\hline
\multirow[t]{5}{*}{$D^*\bar{D}^*\to D^*\bar{D}^*$} 
& $\sigma$ & $1$ & $-1$ & $600$ \\
& $\pi$, $\rho$ & $3/2$ & $-3/2$ & $600$ \\
& $\eta$ & $1/6$ & $-1/6$ & $600$ \\
& $\omega$ & $1/2$ & $-1/2$ & $600$ \\
\multirow[t]{2}{*}{$D^*\bar{D}^*\to D_s\bar{D}_s^*$}
& $K$, $K^*$ & $-\sqrt{2}$ & $-\sqrt{2}$ & $600$ \\
\multirow[t]{2}{*}{$D^*\bar{D}^*\to D_s^*\bar{D}_s^*$}
& $K$, $K^*$ & $-\sqrt{2}$ & $-\sqrt{2}$ & $600$ \\
\hline
\multirow[t]{3}{*}{$D_s\bar{D}_s^*\to D_s\bar{D}_s^*$}
& $\sigma$ & $1$ & - & $600$ \\
& $\eta$ & - & $2/3$ & $600$ \\
& $\phi$ & $1$ & $1$ & $600$ \\
\multirow[t]{3}{*}{$D_s\bar{D}_s^*\to D_s^*\bar{D}_s^*$}
& $\eta$ & $2/3$ & $2/3$ & $600$ \\
& $\phi$ & $1$ & $1$ & $600$ \\
\hline
\multirow[t]{2}{*}{$D_s^*\bar{D}_s^*\to D_s^*\bar{D}_s^*$}
& $\sigma$, $\phi$ & $1$ & $1$ & $600$ \\
& $\eta$ & $2/3$ & $2/3$ & $600$ \\
\hline
\hline
\end{tabular}
\end{table}

\setlength{\tabcolsep}{15pt}
\renewcommand{\arraystretch}{1.5}
\begin{table}[htp]
\centering
\caption{The values of the SU(3) symmetric factor with the channels
containing a charmonium and isospin factor for the kernel amplitude in
isoscalar ($I=0$) channel. $\Lambda_0$ denotes the
reduced cutoff mass defined as $\Lambda-m_\mathrm{ex}$.}
\label{tab:2}
\begin{tabular}{l|c||c|c|c}
\hline
\hline
\multirow{2}{*}{Reaction} & \multirow{2}{*}{Exchange} & 
\multicolumn{2}{c|}{IS factor} & \multirow{2}{*}{$\Lambda_0$ [MeV]} \\
\cline{3-4}
 & & $t$-ch & $u$-ch & \\
\hline
\multirow[t]{2}{*}{$\omega \eta_c\to D\bar{D}$}
& $D^*$ & $-1$ & $-1$ & $600$ \\
\multirow[t]{2}{*}{$\omega \eta_c\to D\bar{D}^*$}
& $D$, $D^*$ & $-1$ & $-1$ & $600$ \\
\multirow[t]{2}{*}{$\omega \eta_c\to D^*\bar{D}^*$}
& $D$, $D^*$ & $-1$ & $-1$ & $600$ \\
\multirow[t]{2}{*}{$\phi \eta_c\to D_s\bar{D}_s$}
& $D_s^*$ & $-\sqrt{2}$ & $-\sqrt{2}$ & $600$ \\
\multirow[t]{2}{*}{$\phi \eta_c\to D_s\bar{D}_s^*$}
& $D_s$, $D_s^*$ & $-\sqrt{2}$ & $-\sqrt{2}$ & $600$ \\
\multirow[t]{2}{*}{$\phi \eta_c\to D_s^*\bar{D_s}^*$}
& $D_s$, $D_s^*$ & $-\sqrt{2}$ & $-\sqrt{2}$ & $600$ \\
\hline
\multirow[t]{2}{*}{$\omega J/\psi\to D\bar{D}$}
& $D$, $D^*$ & $-1$ & $-1$ & $600$ \\
\multirow[t]{2}{*}{$\omega J/\psi\to D\bar{D}^*$}
& $D$, $D^*$ & $-1$ & $-1$ & $600$ \\
\multirow[t]{2}{*}{$\omega J/\psi\to D^*\bar{D}^*$}
& $D$, $D^*$ & $-1$ & $-1$ & $600$ \\
\multirow[t]{2}{*}{$\phi J/\psi\to D_s\bar{D}_s^*$}
& $D_s$, $D_s^*$ & $-\sqrt{2}$ & $-\sqrt{2}$ & $600$ \\
\multirow[t]{2}{*}{$\phi J/\psi\to D_s\bar{D}_s$}
& $D_s$, $D_s^*$ & $-\sqrt{2}$ & $-\sqrt{2}$ & $600$ \\
\multirow[t]{2}{*}{$\phi J/\psi\to D_s^*\bar{D}_s^*$}
& $D_s$, $D_s^*$ & $-\sqrt{2}$ & $-\sqrt{2}$ & $600$ \\
\hline
\hline
\end{tabular}
\end{table}

\section{Results and Discussion}
We are now in a position to present the numerical results and to
discuss them. We focus on the dynamical generation of bound states
and resonances in the transition amplitudes in the hidden-charm sector
over $3.6$ to $4.3\,\mathrm{GeV}$. Since the partial-wave decomposition is most
conveniently performed in the helicity basis using the Wigner
$d$-functions, the transition amplitudes are first obtained by solving
the coupled-channel integral equation~\eqref{eq:5} in the helicity
basis and subsequently transformed to the $LSJ$ basis, which allows us
to analyze resonant behavior with definite spin, orbital, and total
angular momentum. In addition to parity, we impose charge conjugation
and $G$ parity, and classify states by $I^{G}(J^{PC})$. In the
charmonium sector, four families are in principle relevant: $\chi_{cJ}$
with $J^{++}$, $\psi_{J}$ with $J^{--}$, $\eta_{c}$ with $J^{-+}$,
and $h_{c}$ with $J^{+-}$. Since most experimentally reported excited
$c\bar{c}$ states fall into the $\chi_{cJ}$ and $\psi_{J}$ families,
we restrict our analysis to these two sectors. We scan the transition
amplitudes extended into the complex energy plane and search for poles
with the quantum numbers $I^{G}(J^{PC})$ generated dynamically by the
corresponding coupled-channel reactions. The coupling strength to each
hadronic channel is then extracted from the residues of the poles,
providing information on the nature of the bound or resonance states.

\subsection{$\chi_{cJ}$ states: $I^G(J^{PC})=0^+(J^{++})$}
\subsubsection{$\chi_{c0}$ $J^{PC}=0^{++}$}
\begin{figure}[htp] 
\centering
\includegraphics[scale=0.6]{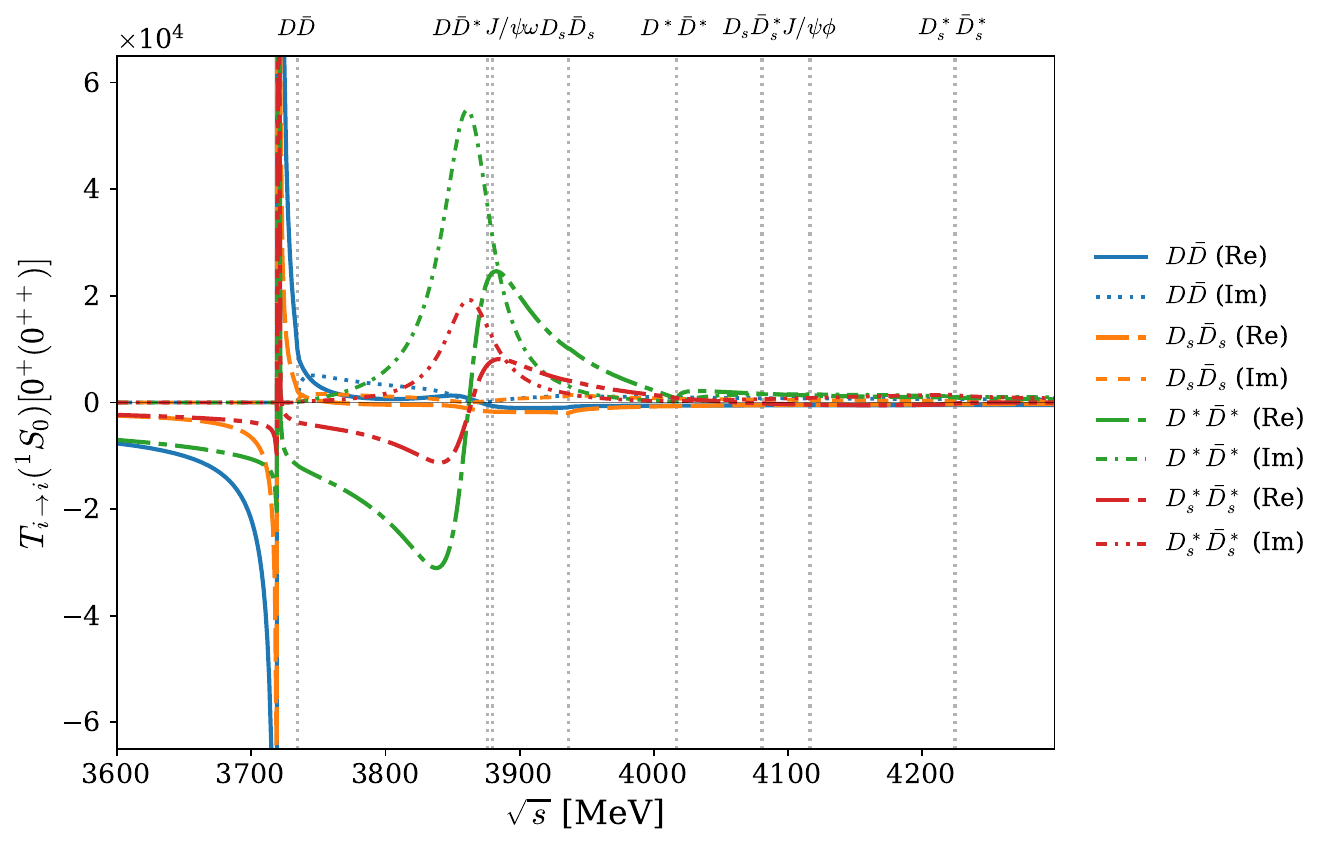}
\caption{Diagonal partial-wave transition amplitudes for the
  $I^G(J^{PC}) = 0^+(0^{++}$) channel as functions of  $\sqrt{s}$.  
}
\label{fig:4}
\end{figure}
Figure~\ref{fig:4} shows the diagonal transition amplitudes for
$\chi_{c0}$ states in the ${}^1S_0$ partial wave near open-flavor
thresholds. Note that charmonium channels do not appear as diagonal
entries in Fig.~\ref{fig:4}, since they do not contribute to elastic
scattering in this partial wave. We find two pole structures in this
channel. The first is a bound state just below the $D\bar{D}$
threshold, generated primarily by elastic $D\bar{D}$ scattering, as
signaled by the sharp variation of the $D\bar{D}$ amplitude near
$\sqrt{s} \approx 3720\,\mathrm{MeV}$ in Fig.~\ref{fig:4}. When all
six hadronic channels listed in Table~\ref{tab:1} are coupled, the
pole is found at $\sqrt{s_R} = 3720.5\,\mathrm{MeV}$, corresponding
to a binding energy of $14\,\mathrm{MeV}$. Experimentally, however,
no bound state with $J^{PC} = 0^{++}$ below the $D\bar{D}$ threshold
has been established. The second is a clear resonance between the
$D\bar{D}$ and $J/\psi\,\omega$ thresholds, with pole position
$\sqrt{s_R} = (3861.34 - i\,22.76)\,\mathrm{MeV}$. As can be seen in 
Fig.~\ref{fig:4}, the most pronounced structure in the transition
amplitude appears in the $D^*\bar{D}^*$ channel near $\sqrt{s}
\approx 3900\,\mathrm{MeV}$, indicating that the dynamical generation
of this resonance is driven primarily by the $D^*\bar{D}^*$ channel,
even though the pole is positioned near the $J/\psi\,\omega$
threshold. We call this pole a $\chi_{c0}(3861)$-like state. Above
the $D_s\bar{D}_s$ threshold we do not find any further peak structure
in this channel. The identification of this state with known
experimental candidates is not straightforward, since there is still
debate over the $\chi_{c0}(2P)$ assignment between $\chi_{c0}(3860)$
and $\chi_{c0}(3915)$, both of which were observed in the $D\bar{D}$
mode. The reported pole positions are
$E_\mathrm{pole}(\chi_{c0}(3860)) = (3862^{+50}_{-35} -
i\,200^{+180}_{-110})\,\mathrm{MeV}$ and
$E_\mathrm{pole}(\chi_{c0}(3915)) = (3922.1\pm1.8 -
i\,20\pm4)\,\mathrm{MeV}$. In mass our pole is closer to
$\chi_{c0}(3860)$, while in width it is closer to $\chi_{c0}(3915)$.

For a quantitative analysis of coupled-channel effects, we extract
the coupling strength to each channel by parametrizing the transition
matrix near $\sqrt{s_R}$ with a Breit-Wigner form,
\begin{align}
  T_{a\to a}(\sqrt{s}) = 4\pi\,\frac{g_a^2}{s - s_R},
  \label{eq:27}
\end{align}
where the residue $g_a$ is the coupling strength for channel $a$.
The results are listed in Table~\ref{tab:3}. The bound state below
the $D\bar{D}$ threshold couples dominantly to $D\bar{D}$, with a
subleading coupling to $D_s\bar{D}_s$, both in $S$ wave. The
resonance at $(3861.34 - i\,22.76)\,\mathrm{MeV}$ couples strongly
to $D^*\bar{D}^*$, and the coupling to $D_s^*\bar{D}_s^*$ is also
large, consistent with strong vector-meson exchange attraction in
$S$ wave. Indeed, this resonance originates as a deep bound state
in the $D^*\bar{D}^*$ channel and is shifted upward by coupled-channel
dynamics, driven predominantly by the $D_s^*\bar{D}_s^*$ channel.

\setlength{\tabcolsep}{15pt}
\renewcommand{\arraystretch}{1.5}
\begin{table}[htp]
\centering
\caption{Coupling strengths $g_i$ for the ${}^{1}S_{0}$ states (in GeV).
Higher partial wave contributions are omitted as they are
negligible. Values in parentheses denote $|g_i|$.} 
\begin{tabular}{l|c|c}
\hline\hline
$\sqrt{s_R}[\text{MeV}]$ & $3720.5$ & $3861.34 - i\,22.76$ \\
\hline
$g_{D\bar D({}^1S_0)}$         & $14.13$  & $5.547 + i\,5.774 \quad (8.007)$ \\
$g_{J/\psi\,\omega({}^1S_0)}$  & $0.190$   & $0.561 - i\,0.099 \quad (0.570)$ \\
$g_{D_s\bar D_s({}^1S_0)}$     & $7.484$   & $2.272 + i\,4.924 \quad (5.423)$ \\
$g_{D^*\bar D^*({}^1S_0)}$     & $2.680$   & $36.41 - i\,16.022 \quad (39.78)$ \\
$g_{J/\psi\,\phi({}^1S_0)}$    & $0.178$   & $0.482 - i\,0.074 \quad (0.488)$ \\
$g_{D_s^*\bar D_s^*({}^1S_0)}$ & $2.228$   & $20.89 - i\,10.769 \quad (23.51)$ \\
\hline\hline
\end{tabular}
\label{tab:3}
\end{table}

\subsubsection{$\chi_{c1}$ $J^{PC}=1^{++}$}
\begin{figure}[htp] 
\centering
\includegraphics[scale=0.6]{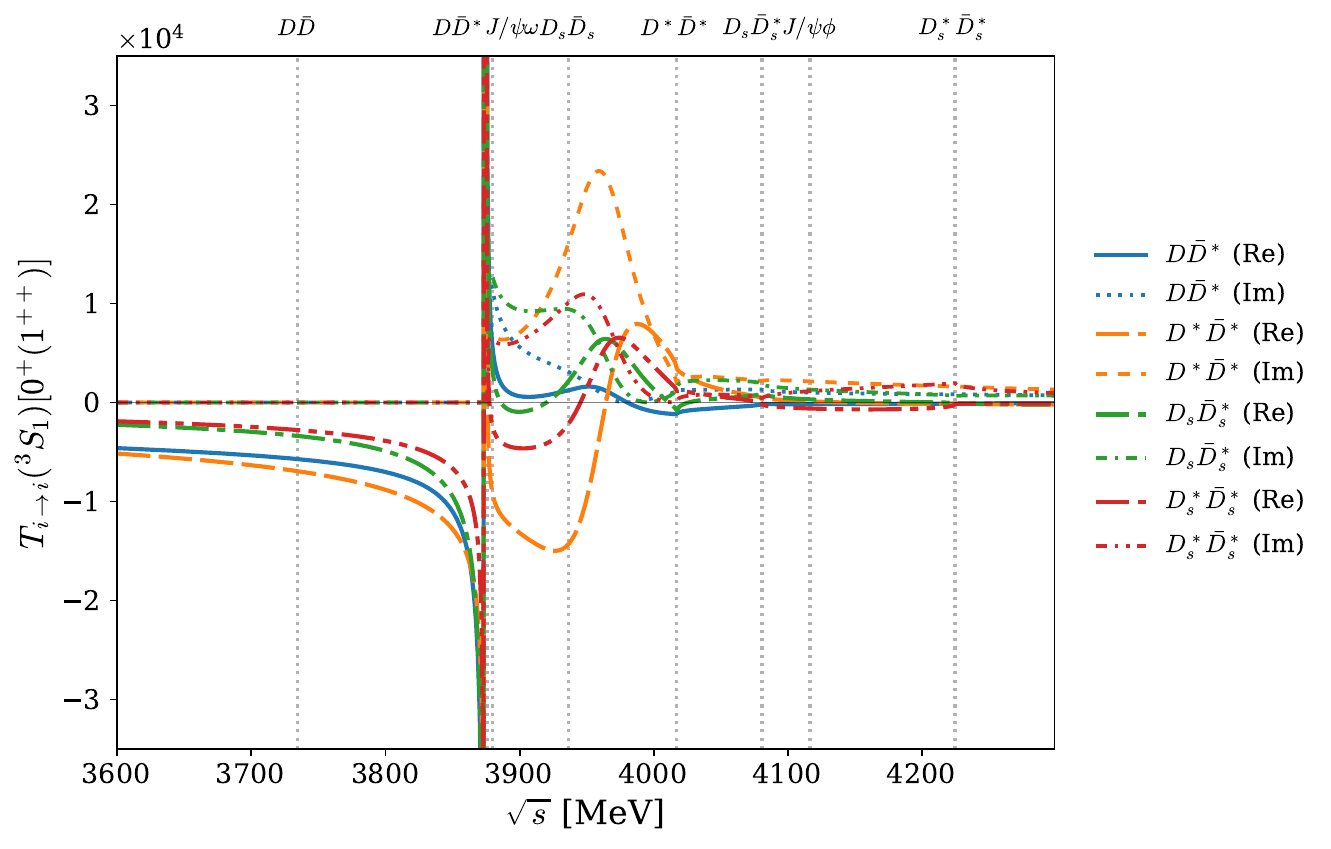}
\caption{
  Axial-vector transition amplitudes for diagonal elements as functions of
  $\sqrt{s}$ for $I^G(J^{PC}) = 0^+(1^{++}$) channel. 
}
\label{fig:5}
\end{figure}
Figure~\ref{fig:5} shows the transition amplitudes for $\chi_{c1}$
states near open-charm thresholds in the $1^{++}({}^3S_1)$ channel.
We find two pole structures in this channel. The first is a bound
state at $\sqrt{s_R} = 3874.26\,\mathrm{MeV}$, lying almost exactly
at the $D\bar{D}^*$ threshold. As can be seen in Fig.~\ref{fig:5},
the sharp variation of the $D\bar{D}^*$ amplitude near this energy
confirms that this state is generated mainly by elastic $D\bar{D}^*$
scattering in $S$ wave, making it a natural candidate for the
well-known $\chi_{c1}(3872)$. The second is a broader resonance
above the $D\bar{D}^*$ threshold at $\sqrt{s_R} = (3961.40 -
i\,32.25)\,\mathrm{MeV}$, whose extracted pole position suggests a
possible identification with the $X(3940)$, observed in the $D\bar
{D}^*$ decay mode but not in $D\bar{D}$~\cite{PDG:2024cfk}.

The channel couplings in Table~\ref{tab:4} clarify the dynamical
origin of the two $1^{++}$ poles. For the bound state at $\sqrt{s_R}
= 3874.26\,\mathrm{MeV}$, the largest coupling is to $D\bar{D}^*$
and $D_s\bar{D}_s^*$ in ${}^3S_1$. Interestingly, the couplings to
$D^*\bar{D}^*({}^3S_1)$ and $D_s^*\bar{D}_s^*({}^3S_1)$ are
comparable to those for $D\bar{D}^*$ and $D_s\bar{D}_s^*$,
indicating a nontrivial structure of the $\chi_{c1}(3872)$ beyond
a pure $D\bar{D}^*$ molecular picture, even without invoking isospin
breaking. Hidden-charm channels are small, and the $D$-wave
components are essentially negligible. For the higher pole at
$\sqrt{s_R} = (3961.40 - i\,32.25)\,\mathrm{MeV}$, the
$S$-wave vector-meson pair channel $D^*\bar{D}^*({}^3S_1)$ carries
the dominant coupling. The $D_s\bar{D}_s^*({}^3S_1)$ and $D\bar
{D}^*({}^3S_1)$ couplings are sizable but subleading, and the
$D$-wave components are smaller by one to two orders of magnitude
than the $S$-wave contributions. Thus this resonance, dominated by
open-flavor $S$-wave channels, is consistent with the broad
structure of the $X(3940)$.
 
\setlength{\tabcolsep}{15pt}
\renewcommand{\arraystretch}{1.5}
\begin{table}[htp]
\centering
\caption{Coupling strengths $g_i$ for the ${}^3S_1$ and ${}^3D_1$
channels (in GeV). For the bound state at 3874.26 MeV, imaginary parts
are omitted due to their negligible magnitude. Values in parentheses
denote $|g_i|$.} 
\begin{tabular}{l|c|c}
\hline\hline
$\sqrt{s_R}[\text{MeV}]$ & $3874.26$ & $3961.40 - i\,32.25$ \\
\hline
$g_{D \bar{D}^* ({}^3S_1)}$       & $7.125$ & $3.712 + i\,6.796 \quad (7.744)$ \\
$g_{J/\psi \omega ({}^3S_1)}$     & $0.102$ & $0.447 + i\,0.060 \quad (0.451)$ \\
$g_{D^* \bar{D}^* ({}^3S_1)}$     & $4.863$ & $22.03 + i\,1.355 \quad (22.07)$ \\
$g_{D_s \bar{D}_s^* ({}^3S_1)}$   & $6.898$ & $10.73 + i\,8.910 \quad (13.95)$ \\
$g_{J/\psi \phi ({}^3S_1)}$       & $0.018$ & $0.023 + i\,0.044 \quad (0.050)$ \\
$g_{D_s^* \bar{D}_s^* ({}^3S_1)}$ & $4.657$ & $14.33 + i\,6.075 \quad (15.56)$ \\
\hline
$g_{D \bar{D}^* ({}^3D_1)}$       & $0.001$ & $0.179 - i\,1.046 \quad (1.061)$ \\
$g_{J/\psi \omega ({}^3D_1)}$     & $0.000$ & $0.023 - i\,0.039 \quad (0.045)$ \\
$g_{D^* \bar{D}^* ({}^3D_1)}$     & $0.084$ & $0.140 + i\,0.093 \quad (0.168)$ \\
$g_{D_s \bar{D}_s^* ({}^3D_1)}$   & $0.161$ & $0.111 + i\,0.158 \quad (0.193)$ \\
$g_{J/\psi \phi ({}^3D_1)}$       & $0.001$ & $0.000 + i\,0.001 \quad (0.001)$ \\
$g_{D_s^* \bar{D}_s^* ({}^3D_1)}$ & $0.193$ & $0.415 + i\,0.241 \quad (0.480)$ \\
\hline\hline
\end{tabular}
\label{tab:4}
\end{table}

\subsubsection{$\chi_{c2}$ $J^{PC}=2^{++}$}
\begin{figure}[htp] 
\centering
\includegraphics[scale=0.6]{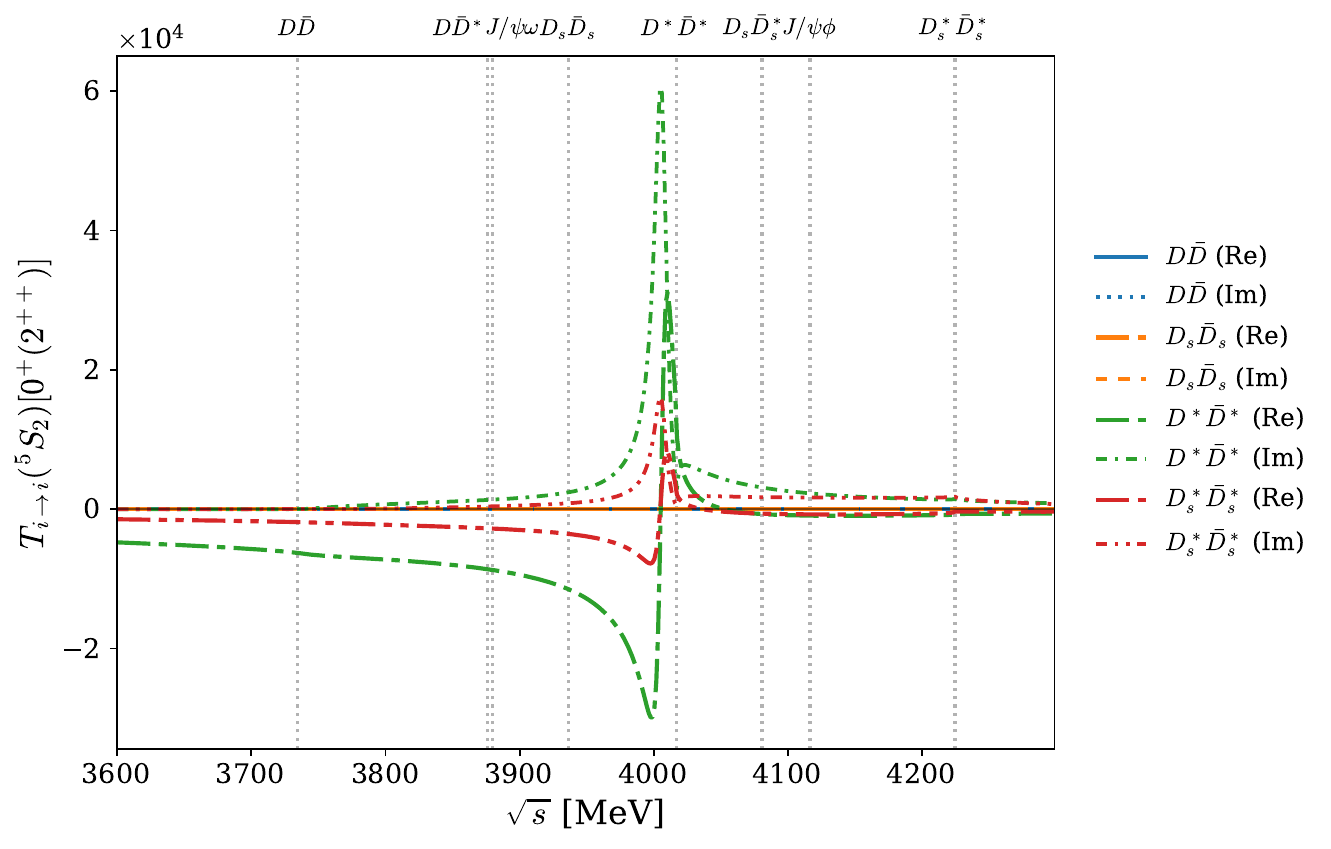}
\includegraphics[scale=0.6]{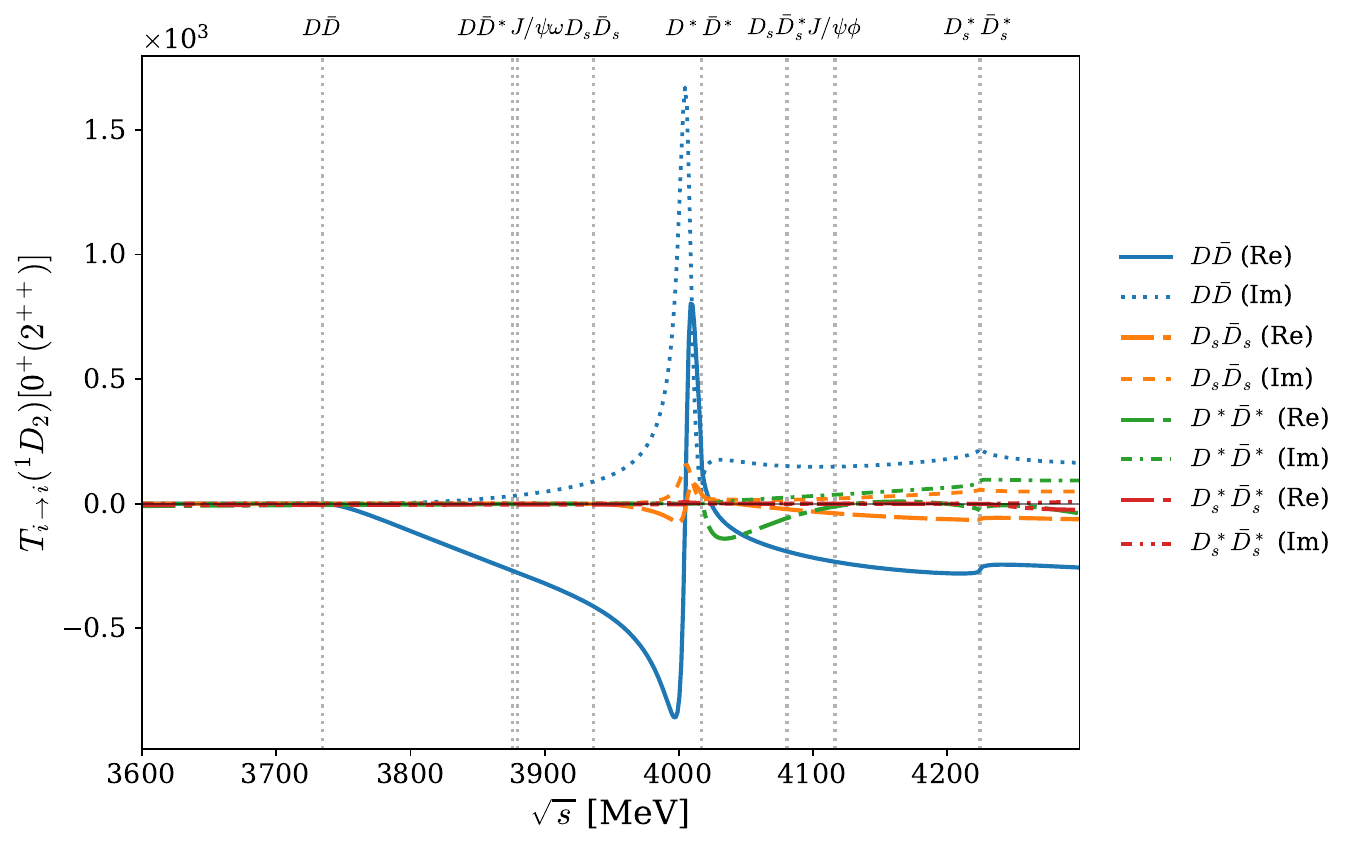}
\caption{
  Tensor transition amplitudes for diagonal elements as functions of
  $\sqrt{s}$ for $I^G(J^{PC}) = 0^+(2^{++}$) channel. 
}
\label{fig:6}
\end{figure}
Figure~\ref{fig:6} shows the transition amplitudes for $\chi_{c2}$
states near open-charm thresholds in the $2^{++}$ channel. We find
a narrow pole at $\sqrt{s_R} = (4005.26 - i\,5.95)\,\mathrm{MeV}$,
which is located between the $D_s\bar{D}_s$ and $D^*\bar{D}^*$
thresholds. This pole appears in both ${}^5S_2$ and ${}^1D_2$ partial
waves, and the small imaginary part indicates a narrow width of about
$12\,\mathrm{MeV}$, suggesting only moderate coupled-channel
effects. The residues in Table~\ref{tab:5} clarify the structure of
this state. The largest coupling is to $D^*\bar{D}^*$ in $S$ wave with
$|g_{D^*\bar{D}^*}| \simeq 1.49$, and this pole also appears in the
$D^*\bar{D}^*$ single-channel solution. As in the $\chi_{c0}$ and
$\chi_{c1}$ cases, the $S$-wave contribution from $D_s^*\bar{D}_s^*$
is comparable to the dominant one, with $|g_{D_s^*\bar{D}_s^*}| 
\simeq 0.88$. The hidden-charm channel couplings are negligibly
small. Among the $D$-wave contributions, the $D\bar{D}({}^1D_2)$
component is the largest with $|g_{D\bar{D}}| \simeq 0.28$,
consistent with the peak structure seen in the lower panel of
Fig.~\ref{fig:6}.
We note that the pole mass lies about $70\,\mathrm{MeV}$ above the
typical values quoted for $\chi_{c2}(3930)$. Whether this state
corresponds to a new tensor charmonium not yet observed experimentally
remains to be seen.

\setlength{\tabcolsep}{15pt}
\renewcommand{\arraystretch}{1.5}
\begin{table}[htp]
\centering
\caption{Coupling strengths $g_i$ for the ${}^5S_2$ and ${}^1D_2$
channels (in GeV). For the lower pole, imaginary parts are omitted
when negligible. Unphysical or zero-coupling states are omitted for brevity.}
\begin{tabular}{l|c|c}
\hline\hline
 & $g_i$ & $|g_i|$ \\
\hline
$g_{J/\psi \omega({}^5S_2)}$            & $0.003 - i\,0.002$   & $0.004$ \\
$g_{D^* \bar D^*({}^5S_2)}$             & $1.692 - i\,0.216$   & $1.706$ \\
$g_{J/\psi \phi({}^5S_2)}$              & $0.003 + i\,0.000$   & $0.003$ \\
$g_{D_s^*\bar D_s^*({}^5S_2)}$          & $0.877 - i\,0.066$   & $0.879$ \\
\hline
$g_{D\bar{D}({}^1D_2)}$                 & $0.283 - i\,0.015$   & $0.283$ \\
$g_{D_s \bar D_s({}^1D_2)}$             & $0.086 - i\,0.018$   & $0.088$ \\
$g_{D^* \bar D^*({}^1D_2)}$             & $0.002 + i\,0.001$   & $0.002$ \\
$g_{D_s^*\bar D_s^*({}^1D_2)}$          & $0.019 - i\,0.001$   & $0.019$ \\
\hline\hline
\end{tabular}
\label{tab:5}
\end{table}

\subsection{$\psi$ states: $I^G(J^{PC})=0^-(J^{--})$}
Using the same coupled-channel framework, we analyze the transition
amplitudes for the vector charmonium $\psi$ states with several
values of $J$. Within the energy range of interest, we do not find
dynamically generated states for $J=0$, $J=1$, or $J=2$. Based on
quark models, the well-known states in these sectors, such as
$\psi(4040)$ with $J^{PC}=1^{--}$, can be regarded as having a
dominant $c\bar{c}$ core, suggesting that their description would
require the inclusion of $s$-channel pole diagrams, which have been
intentionally excluded in the present framework. As the current
approach is designed to capture solely the states generated by
coupled-channel dynamics, the description of these sectors could be
improved by incorporating $s$-channel contributions, which we leave
for future work. In contrast, for $J=3$ we find a pole above the
$D^*\bar{D}^*$ threshold in the $J^{PC}=3^{--}$ sector, which is
less likely to be dominated by a $c\bar{c}$ core and is therefore
a more natural candidate for a dynamically generated state. We
discuss this state in detail below.

\subsubsection{$\psi_3$ $J^{PC}=3^{--}$}
\begin{figure}[htp] 
\centering
\includegraphics[scale=0.6]{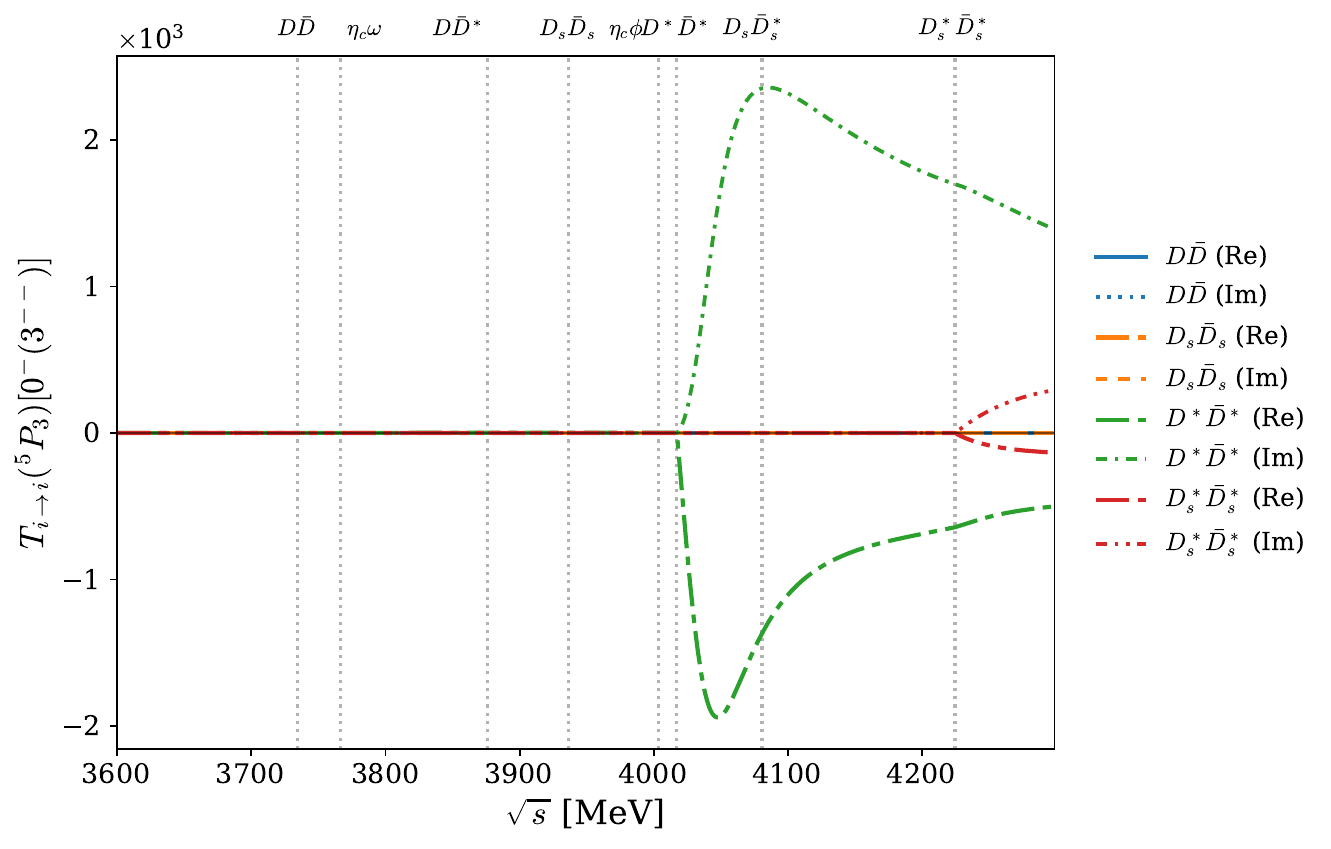}
\includegraphics[scale=0.6]{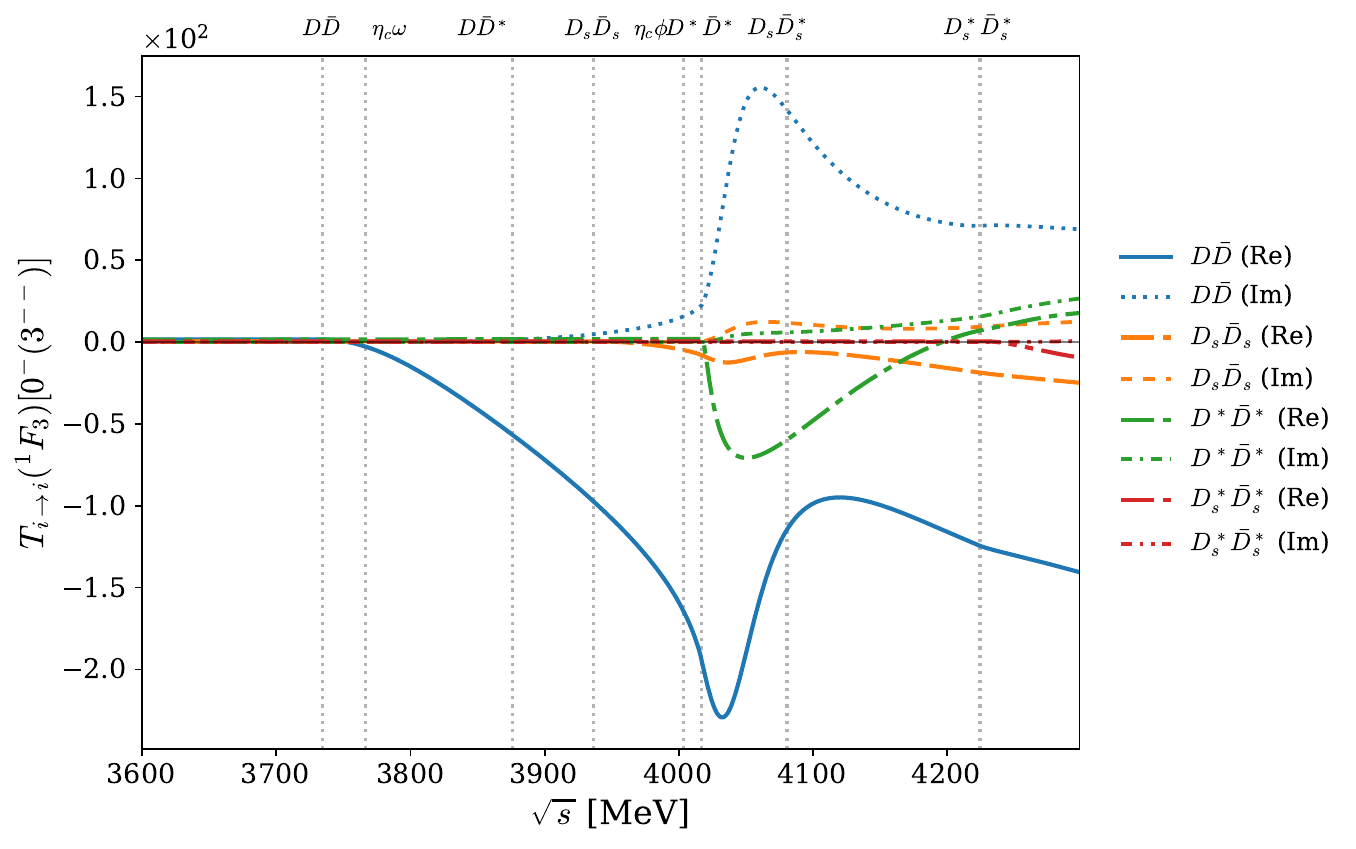}
\caption{
  Transition amplitudes for diagonal elements as functions of
  $\sqrt{s}$ for $I^G(J^{PC}) = 0^-(3^{--}$) channel in ${}^5P_3$ and
  ${}^1F_3$ partial waves.}
\label{fig:7}
\end{figure}
For the $\psi_3$ state with $J^{PC}=3^{--}$, the selection rules
require odd orbital angular momentum $L$ and, for identical meson
pairs, even total spin $S$. Consequently, the allowed partial waves
are restricted to ${}^1F_3$ for $D\bar{D}$ and $D_s\bar{D}_s$, and
${}^1F_3$, ${}^5P_3$, and ${}^5F_3$ for $D^*\bar{D}^*$ and
$D_s^*\bar{D}_s^*$. Non-identical pairs and hidden-charm channels
couple through the ${}^3F_3$ state. Figure~\ref{fig:7} shows the
diagonal transition amplitudes near open-charm thresholds over
$3.6$--$4.3\,\mathrm{GeV}$ for $J^{PC}=3^{--}$ in the dominant
${}^5P_3$ and ${}^1F_3$ partial waves. In the $P$-wave transition
amplitude shown in the top panel of Fig.~\ref{fig:7}, we find a
rapid threshold enhancement of the elastic $D^*\bar{D}^*$ scattering
just above the $D^*\bar{D}^*$ threshold. The elastic $D\bar{D}$
transition amplitude in $F$ wave also exhibits a clear peak structure
just above the $D^*\bar{D}^*$ threshold. Scanning the complex energy
plane of the transition matrix, we locate a pole at $\sqrt{s_R} =
(4030.48 - i\,33.33)\,\mathrm{MeV}$.

The channel coupling strengths listed in Table~\ref{tab:6} confirm
the dynamical picture suggested by the line shapes. The dominant
coupling is to the $D^*\bar{D}^*({}^5P_3)$ channel with
$|g_{D^*\bar{D}^*}| \simeq 7.663$, consistent with the pronounced
threshold enhancement seen in the top panel of Fig.~\ref{fig:7}.
The $D\bar{D}({}^1F_3)$ coupling is also sizable with
$|g_{D\bar{D}}| \simeq 2.057$, in accordance with the peak
structure in the lower panel. The remaining partial-wave
contributions are considerably smaller. This pole lies in the
vicinity of the experimentally known $\psi_3(3842)$ with
$J^{PC}=3^{--}$~\cite{PDG:2024cfk}, though its mass is about
$190\,\mathrm{MeV}$ higher, and we therefore regard it as a
candidate for a new $3^{--}$ charmonium state.

\setlength{\tabcolsep}{12pt}
\renewcommand{\arraystretch}{1.4}
\begin{table}[htp]
\centering
\caption{Non-zero coupling strengths $g_i$ for the allowed partial
  waves of the $\psi_3$ state (in GeV). Unphysical partial waves
  forbidden by spin, parity, and C-parity conservation are omitted.} 
\begin{tabular}{l|c|c}
\hline\hline
 & $g_i$ & $|g_i|$ \\
\hline
$g_{D^* \bar D^*({}^5P_3)}$      & $2.883 - i\,7.100$ & $7.663$ \\
$g_{D_s^*\bar D_s^*({}^5P_3)}$   & $0.004 - i\,0.002$ & $0.004$ \\
\hline
$g_{D\bar{D}({}^1F_3)}$          & $1.326 - i\,1.573$ & $2.057$ \\
$g_{D_s \bar D_s({}^1F_3)}$      & $0.351 - i\,0.486$ & $0.599$ \\
$g_{D^* \bar D^*({}^1F_3)}$      & $0.289 - i\,0.222$ & $0.364$ \\
$g_{D_s^*\bar D_s^*({}^1F_3)}$   & $2.661\times 10^{-4} - i\,0.001$ & $1.228\times 10^{-3}$ \\
\hline
$g_{D\bar{D}^*({}^3F_3)}$        & $0.008 + i\,9.985\times 10^{-5}$ & $0.008$ \\
$g_{\eta_c \omega({}^3F_3)}$     & $1.874\times 10^{-4} - i\,8.331\times 10^{-5}$ & $2.051\times 10^{-4}$ \\
$g_{D_s \bar D_s^*({}^3F_3)}$    & $2.152\times 10^{-6} + i\,5.338\times 10^{-4}$ & $5.338\times 10^{-4}$ \\
$g_{\eta_c \phi({}^3F_3)}$       & $0$ & $0$ \\
\hline
$g_{D^* \bar D^*({}^5F_3)}$      & $0.702 - i\,0.245$ & $0.744$ \\
$g_{D_s^*\bar D_s^*({}^5F_3)}$   & $8.869\times 10^{-4} - i\,0.002$ & $0.002$ \\
\hline\hline
\end{tabular}
\label{tab:6}
\end{table}

\section{Summary and conclusions}
In the present work, we have investigated the dynamical generation
of charmonium-like states with isospin $I=0$ and spin-parity
$J^{PC} = 0^{++}$, $1^{++}$, $2^{++}$, and $3^{--}$ in the
mass range of $3.6$ to $4.3\,\mathrm{GeV}$, employing an off-shell
coupled-channel formalism based on the Blankenbecler-Sugar
reduction of the Bethe-Salpeter equation. The kernel amplitudes
were constructed from effective Lagrangians respecting heavy-quark
spin-flavor and chiral symmetries, including $t$- and $u$-channel
meson exchanges while intentionally excluding $s$-channel pole
diagrams, so that all identified poles are generated solely by
coupled-channel dynamics. Having solved the coupled-channel integral
equations and scanned the transition amplitudes in the complex energy
plane, we have identified six poles in total. Their properties are
summarized as follows: 
\begin{itemize}
\item In the scalar ($0^{++}$) sector, two poles are identified.
The first is a bound state at $\sqrt{s_R} = 3720.5\,\mathrm{MeV}$,
lying $14\,\mathrm{MeV}$ below the $D\bar{D}$ threshold and
generated primarily by elastic $D\bar{D}$ scattering; no
experimental evidence has been established for this state.
The second pole at $\sqrt{s_R} = (3861.34 - i\,22.76)\,\mathrm{MeV}$
is dynamically generated by the $D^*\bar{D}^*$ channel despite
being positioned near the $J/\psi\,\omega$ threshold; it is
consistent in mass with $\chi_{c0}(3860)$ and in width with
$\chi_{c0}(3915)$, and the assignment to either candidate remains
unclear. 
\item
In the axial-vector ($1^{++}$) sector, we reproduce a narrow bound
state at $\sqrt{s_R} = 3874.26\,\mathrm{MeV}$, lying almost exactly
at the $D\bar{D}^*$ threshold and generated predominantly by elastic
$D\bar{D}^*$ scattering. This state is a natural candidate for the
well-known $\chi_{c1}(3872)$. The coupling analysis reveals a
nontrivial channel structure beyond a pure $D\bar{D}^*$ molecular
picture, with sizable contributions from $D^*\bar{D}^*$ and
$D_s\bar{D}_s^*$. A second, broader pole is found at $\sqrt{s_R} =
(3961.40 - i\,32.25)\,\mathrm{MeV}$, dominated by the
$D^*\bar{D}^*({}^3S_1)$ channel, and is identified as a plausible
candidate for the $X(3940)$.
\item
In the tensor ($2^{++}$) sector, we find a narrow pole at
$\sqrt{s_R} = (4005.26 - i\,5.95)\,\mathrm{MeV}$, lying below the
$D^*\bar{D}^*$ threshold and appearing in both ${}^5S_2$ and
${}^1D_2$ partial waves. The dominant coupling is to $D^*\bar{D}^*$
in $S$ wave, and the pole also appears in the $D^*\bar{D}^*$
single-channel, indicating that coupled-channel effects
are moderate. The pole mass lies about $70\,\mathrm{MeV}$ above
the typical values quoted for $\chi_{c2}(3930)$.
This state has not yet been observed experimentally.
\item
In the vector ($3^{--}$) sector, we find a pole at $\sqrt{s_R} =
(4030.48 - i\,33.33)\,\mathrm{MeV}$, above the $D^*\bar{D}^*$
threshold. This state is dominantly coupled to $D^*\bar{D}^*$ in
the ${}^5P_3$ partial wave, with a sizable $D\bar{D}({}^1F_3)$
contribution. We regard this as a candidate for a new $3^{--}$
charmonium state, distinct from $\psi_3(3842)$, whose mass lies
about $190\,\mathrm{MeV}$ lower.
\end{itemize}

A common pattern emerging from the present analysis is the crucial
role of the $D^*\bar{D}^*$ channel in driving the dynamical
generation of the resonances across all spin-parity sectors
investigated. In particular, the $D_s^*\bar{D}_s^*$ channel
consistently provides a subleading but nonnegligible contribution,
reflecting the importance of hidden-strangeness components in the
coupled-channel dynamics.
We note that no dynamically generated poles are found in the $\psi$
sectors with $J = 0$, $1$, and $2$. This may be related to the fact
that the well-known states in these sectors, such as $\psi(4040)$,
are widely regarded as having a dominant $c\bar{c}$ core, whose
description would require the inclusion of $s$-channel contributions
not considered in the present framework. A more complete treatment
incorporating both $s$-channel pole diagrams and coupled-channel
dynamics would be needed to describe the full charmonium spectrum,
and we leave this for future work.
\begin{acknowledgments}
The present work was supported by the National Research Foundation of
Korea (NRF) grant funded by the Korea government under Grant
No. RS-2025-02634319 (HJK) and RS-2025-00513982 (HChK).
\end{acknowledgments}

\bibliography{chi_c0}
\bibliographystyle{apsrev4-1}

\end{document}